\documentclass[fleqn,usenatbib]{mnras}

\usepackage{newtxtext,newtxmath}
\usepackage{xcolor}
\usepackage{booktabs}
\usepackage{adjustbox}
\usepackage[normalem]{ulem}

\defcitealias{LimaNeto25}{Paper~I}

\usepackage[T1]{fontenc}

\DeclareRobustCommand{\VAN}[3]{#2}
\let\VANthebibliography\thebibliography
\def\thebibliography{\DeclareRobustCommand{\VAN}[3]{##3}\VANthebibliography}

\title[Simulations of NGC~5098 group]{Simulations of collision and sloshing in the galaxy group NGC 5098/5096}

\author[Albuquerque et al.]{%
Richards P. Albuquerque$^{1}$\thanks{E-mail: albuquerque.richards.astro@gmail.com},
Gastão B. Lima Neto$^{1}$,
Rubens E. G. Machado$^{2}$, 
Hugo V. Capelato$^{3}$,
\and Florence Durret$^{4}$
\\
$^{1}$Instituto de Astronomia, Geof\'isica e Ci\^encias Atmosf\'ericas, Universidade de S\~ao Paulo, Rua do Mat\~ao 1226, S\~ao Paulo, SP, Brazil\\
$^{2}$Departamento Acad\^emico de F\'isica, Universidade Tecnol\'ogica Federal do Paran\'a, Av. Sete de Setembro 3165, Curitiba, PR, Brazil \\
$^{3}$N\'ucleo de Astrof\'\i sica Te\'orica (NAT), Universidade Cidade de S\~ao Paulo, Rua Galv\~ao Bueno, 868, 01506-000, São Paulo, Brazil\\
$^{4}$Sorbonne Universit\'e, CNRS, UMR~7095, Institut d'Astrophysique de Paris, 98bis Bd Arago, 75014, Paris, France
\\
}

\date{Accepted 202X Month XX. Received 202X Month XX; in original form 202X Month XX}

\pubyear{2025}

\begin{document}
\label{firstpage}
\pagerange{\pageref{firstpage}--\pageref{lastpage}}
\maketitle

\begin{abstract}
The study of galaxy groups is essential to understanding the evolutionary history of large-scale structures in the Universe. These dense environments have a significant impact on galaxy evolution, influencing their gas content, morphology, and star formation activity. In this work we analyse in detail the system NGC~5098$/$5096 composed of two galaxy groups. We performed hydrodynamical $N$-body simulations of a galaxy group collision aimed at reproducing the gas sloshing and surface brightness distribution observed in X-ray data. We conducted a detailed X-ray analysis and generated mock image \textit{Chandra} observations from our simulations. The resulting corrected mock image surface brightness profiles show good agreement with the observed data. The relative line-of-sight velocity between NGC~5098 and NGC~5096 is $v_{\mathrm{los}} = 700$ km s$^{-1}$, with a projected separation of $d_{\mathrm{proj}} = 155$ kpc, suggesting that the collision occurs nearly in the line-of-sight. Our simulations were performed with an inclination angle of $80^\circ$ in order to reproduce the dynamical constraints. We also find a correlation between the dark matter and intragroup light distributions when comparing the residual dark matter map with the intragroup light morphology. Our best-fitting model is consistent with these observational constraints and provides a plausible dynamical scenario for the current state of the NGC~5098 group interaction with NGC 5096.
\end{abstract}




\section{Introduction}

Galaxy groups are gravitationally bound systems that contain more than half of all galaxies in the low redshift Universe \citep[e.g.][]{Eke04,Robotham11}. These systems arise naturally in a Universe where structure forms hierarchically, with the merging of smaller groups leading to increasingly massive halos, ultimately forming rich galaxy clusters \citep[e.g.][]{Press74,Fakhouri10}. Galaxy groups occupy an intermediate mass regime between massive elliptical galaxies ($\sim 10^{12}$ M$_\odot$) and galaxy clusters ($\sim 10^{14}$ M$_\odot$), and they host a substantial fraction of the baryonic matter in the local Universe \citep[e.g.][]{Fukugita98,Eke04}.

In terms of their optical and physical properties, galaxy groups have been broadly classified into different categories: poor (or loose), compact, rich, and fossil groups \citep[e.g.][]{Lovisari21}. Dynamically, compact groups, being relatively isolated by definition \citep{Hickson82}, tend to exhibit more relaxed configurations. Conversely, many galaxy groups are dynamically unrelaxed, undergoing mergers and interactions with infalling substructures \citep[e.g.][]{Tempel17, Ewan19}. 

In the X-ray band, many groups show extended diffuse emission of the intragroup medium (hereafter IGM), with features such as surface brightness discontinuities, cold fronts, and sloshing \citep[e.g.][]{Markevitch07, Lagana2010, Gastaldello13}.

The \textit{Chandra} and XMM-\textit{Newton} X-ray observatories have enabled detailed investigations of collision and merger dynamics in galaxy groups and clusters. By combining these observations with $N$-body hydrodynamical simulations we have better understood the origin of shock fronts \citep[e.g.][]{Springel07, Donnert14, Moura21, Albuquerque24,  Chadayammuri24, Machado24} and cold fronts \citep[e.g.][]{Markevitch07,Roediger12, Gastaldello13, Doubrawa22, Machado22} in the intragroup and intracluster medium.

Galaxy groups and clusters often host active galactic nuclei (hereafter AGN) with jets of relativistic plasma that emit synchrotron radiation \citep[e.g.][]{Tadhunter16}. The mechanical feedback from AGN is an important mechanism for heating the gas in cool-core regions. Evidence of this mechanical feedback is found in the X-ray cavities observed in galaxy groups and clusters, on scales roughly coincident with the lobes of the central radio galaxy \citep[e.g.][]{McNamara00}. The radio plasma emitted by the AGN displaces the gas, generating low-density bubbles that distribute energy to the surrounding medium. Such cavities or bubbles, identified by depressions in surface brightness, can extend to distances of up to 186\,kpc, as observed in the extreme case of the galaxy cluster MS 0735.6+7421 \citep{Biava21}.

Galaxy groups have shallower potential wells and thus lower mean temperatures than clusters. Therefore non-gravitational processes such as galactic winds, radiative cooling, and AGN feedback, play a more significant role in their evolution. In this environment, even a single massive galaxy falling into a group-scale halo with a nonzero impact parameter can perturb the IGM, inducing gas motions, transferring angular momentum, and producing a spiral pattern in the IGM that also manifests as sloshing features.

X-ray analyses of interacting galaxy groups \citep[e.g.][]{Xue04, Kraft06, Randall09, Machacek10, Gastaldello13, Sullivan14, Schellenberger23, LimaNeto25} are relatively less common than interacting clusters due to their intrinsically lower luminosities. Since galaxy groups outnumber clusters \citep{Tully87, Eke04}, understanding the range of collision and merger-driven effects on their gas and galaxy properties is essential for constructing accurate models of hierarchical structure formation. 

NGC~5098 is a galaxy group dominated by a central pair of elliptical galaxies, NGC~5098a and NGC~5098b, identified by \citet{Ramella95}. Based on the SDSS\footnote{\url{https://skyserver.sdss.org/dr18}} data, the brightest galaxy is NGC~5098a, with absolute magnitudes of $M_B = -21.131$ and $M_V = -22.097$, as computed by \cite{Randall09}, and redshift $z = 0.03606 \pm 0.00001$ \citep{Lee17}. NGC~5098b has absolute magnitudes of $M_B = -20.845$ and $M_V = -21.770$, with redshift $z = 0.03744 \pm 0.00008$ \citep{Miller02}. Their line-of-sight relative velocity is $v_\mathrm{los} = 413$\,km\,s$^{-1}$, and their projected separation is $d_\mathrm{proj} = 27$\,kpc. Note that the western galaxy, NGC~5098a, hosts the extended radio source B2\,1317+33, which has been detected at 1415, 4850, and 4996\,MHz \citep{Fanti77, Becker91}.

At a projected distance of 155\,kpc southwest of NGC~5098 lies the galaxy group NGC~5096 originally considered part of NGC~5098 \citep[e.g.][]{Ramella95} but already recognized as an independent sub-system by  \cite{Mahdavi05}. 

As discussed by \cite{LimaNeto25} \citepalias[hereafter][]{LimaNeto25}, the overall structure of NGC~5098 comprises two subsystems: i) a dominant group with mean redshift $z=0.03605$ centred on NGC~5098a and having virial mass of $M_{200} = (3.93 \pm 0.28) \times 10^{13}\,\mathrm{M}_\odot$ and radius $R_{200} = (0.69 \pm 0.02)$\,Mpc. This group comprises 82 galaxies to $r <18$~mag; ii) a smaller group with mean redshift $z = 0.03886$ distributed around the compact group NGC~5096 (\cite{Zandivarez22}) with virial mass $M_{200} = (0.17 \pm 0.02) \times 10^{13}\,\mathrm{M}_\odot$  and radius $R_{200} = (0.24 \pm 0.01)\,\mathrm{Mpc}$ \citepalias{LimaNeto25}.  This implies a mass ratio of approximately 1:20 for the NGC~5098/5096 system. This group has about 30 members ($r <18$~mag) from which five belong to the compact group itself.

NGC~5098 was detected as an extended X-ray source in the ROSAT All-Sky Survey \citep[RASS,][]{Mahdavi00}. Subsequent studies of its global structure have been carried out with XMM-\textit{Newton} \citep{Xue04} and, with higher spatial resolution, with \textit{Chandra} \citep{Sun09}, while \citet{Randall09} used deep \textit{Chandra} data obtained with the more sensitive detector ACIS-S3. They identified a spiral-like arm that can be interpreted as a sloshing pattern originating at NGC~5098a and winding outward in a clockwise direction from northeast to west. They also reported X-ray cavities coincident with VLA 1.45\,GHz radio emission, indicating recent AGN-driven outbursts.

Analysis of the deep \textit{Chandra} imaging by \cite{Randall09} shows that there are 14 cavities detected within a radius of 22\,kpc centred on the AGN, with an average cavity volume of 31\,kpc$^3$. Two of these cavities, located to the north and southeast of the central AGN, are spatially correlated with radio emission and, based on pressure estimates and their proximity to the AGN, are likely recent features. The other X-ray cavities, which show no correlation with radio emission, appear to be ghost cavities from previous AGN outbursts \citep{Randall09}.

In galaxy groups and clusters there is a low surface brightness stellar component, typically found in the surroundings of the brightest cluster galaxies and massive satellites. This diffuse light originates from stars stripped from galaxies through tidal interactions \citep{Contini21, Montes22}. In this work, we refer to this component as the intracluster light (hereafter ICL). Recently our team conducted a detailed study of the ICL in NGC~5098 \citetalias{LimaNeto25}, in which we detected low surface brightness diffuse emission tracing past galaxy-galaxy interactions and also suggesting a past collision between the two galaxy groups, a hypothesis that was already suggested by the kinematical analysis that demonstrated the reality of the NGC~5096 substructure.

In the present work, we aim to model through hydrodynamical $N$-body simulations the dynamical history of the possible interaction between the NGC~5098 and NGC~5096 groups. We also investigate the excesses in the dark matter distribution and compare them with our recent results on the ICL distribution \citetalias{LimaNeto25}, in order to test the recent suggestion \citep[e.g.][]{Montes19, Kluge21} that the ICL component traces the properties of its host system, making it a potential tracer of the dark matter.

This paper is organized as follows. In Section~\ref{method}, we describe the simulation setup, including the initial conditions and the range of dynamical parameters explored. Section~\ref{results} presents the results for our best‐fit scenario, beginning with a comparison between the observed and corrected mock \textit{Chandra} emissivity maps, followed by a discussion of how variations in the initial conditions affect the sloshing signature. Finally, Section~\ref{conclusions} summarizes our findings and provides concluding remarks. We assume a standard $\Lambda$CDM cosmology with $\Omega_{\Lambda} = 0.73$, $\Omega_{\mathrm{m}} = 0.27$, and $H_{0} = 70$\,km\,s$^{-1}$\,Mpc$^{-1}$.

\section{Simulation setup}
\label{method}

In order to model the gas sloshing in NGC~5098 with tailored hydrodynamic $N$-body simulations, we employed \textsc{Gadget-4} \citep{Springel21}. This galaxy group exhibits sloshing morphology and edge-like features in both the northeastern and southwestern regions \citep{Randall09}, indicative of a past dynamical interaction. Therefore, a successful model must reproduce the observed sloshing signatures and recover the locations of these edge-like discontinuities.

\subsection{Initial conditions}

To model the initial conditions of both galaxy groups, we adopt analytical density profiles that are widely used in the literature to describe the dark matter and gas components in equilibrium. For the dark matter halo, we assume a \citet{Hernquist90} profile:
\begin{equation}
\rho_{\mathrm{h}}(r) = \frac{M_{\mathrm{h}}}{2\pi} \frac{a_{\mathrm{h}}}{r(r + a_{\mathrm{h}})^3} ~,
\end{equation}
where $ M_{\mathrm{h}} $ is the total dark matter mass and $ a_{\mathrm{h}} $ is the scale length. The corresponding cumulative mass profile is given by:
\begin{equation}
M(<r) = M_{\mathrm{h}} \frac{r^2}{(r + a_{\mathrm{h}})^2} ~.
\end{equation}
This profile is particularly suitable for our simulations for several reasons. In the inner regions, its shape closely resembles that of the \citet{NFW96} (hereafter NFW) profile, which is widely used to model dark matter halos. However, the Hernquist profile features a steeper decline at large radii, resulting in a finite total mass. This property makes it especially advantageous for modelling isolated systems, as it eliminates the need for an artificial truncation of the halo. 

Given that both profiles are similar at small radii ($ R \ll R_{200} $), we can approximate the Hernquist profile by the NFW profile in this regime, i.e., $ \rho_{\mathrm{h}} \approx \rho_{\mathrm{NFW}} $ \citep{Springel05}. This approximation implies a correspondence between the Hernquist scale length $ a_{\mathrm{h}} $ and the NFW scale radius $ r_{\mathrm{s}} $, where the latter is defined through the concentration parameter $ c $. To calculate the scale length for the Hernquist density profile, we used the following relation:
\begin{equation}
    a_\mathrm{h} = r_\mathrm{s} \sqrt{2\left[ \ln(1 + c) - \frac{c}{1+c} \right]} ~,
\end{equation}
where $ r_\mathrm{s} $ is the scale radius of the NFW density profile. Using the concentration definition, we have $ r_\mathrm{s} = R_{200}/c $. The concentration parameter $ c $ was estimated using the mass-concentration relation from \citet{Duffy08}:
\begin{equation}
    c = \frac{6.71}{\left( 1 + z \right)^{0.44}} \left( \frac{M_{200}}{2 \times 10^{12} \, h^{-1} \, \mathrm{M_\odot}} \right)^{-0.091} ~,
\end{equation}
which depends only on the halo mass and redshift of the galaxy group.

To model the density distribution of the IGM in our simulated galaxy groups collision, we adopted the analytical profile proposed by \citet{Dehnen93}:
\begin{equation}
\rho_{\mathrm{g}}(r) = \frac{(3 - \gamma) M_{\mathrm{g}}}{4\pi} \frac{a_{\mathrm{g}}}{r^{\gamma}(r + a_{\mathrm{g}})^{4 - \gamma}} ~,
\end{equation}
where $M_{\mathrm{g}}$ is the total gas mass, $a_{\mathrm{g}}$ is the gas scale length, and $\gamma$ describes the inner power-law slope of the profile. When $\gamma = 0$, the gas distribution exhibits a flat core (i.e., constant central density), whereas when $\gamma = 1$ the gas distribution is similar to the Hernquist profile, characterized by a steep central density. In our simulations, we adopted $\gamma = 1$ for NGC~5098, representing a cool-core galaxy group, and $\gamma = 0$ for NGC~5096, corresponding to a non-cool-core system. This choice is consistent with previous X-ray studies \citep[e.g.,][]{Xue04, Randall09}.

For the initial condition, we assumed a gas fraction of $f_\mathrm{gas} = 0.1$, a typical value consistent with the mass range of galaxy groups \citep{Lagana13}. Additionally, to determine the gas scale length $ a_{\mathrm{g}} $, we used the total gas mass and the central gas density, which is defined as $\rho_0 = \mu_{\mathrm{e}} n_{\mathrm{e}} m_\mathrm{H}$, where $\mu_{\mathrm{e}} = 1 / ( X + \frac{Y}{2} )$ is the mean molecular weight per electron. Assuming a primordial abundance with hydrogen and helium mass fractions $X = 0.76$ and $Y = 0.24$, we obtain $\mu_{\mathrm{e}} \approx 1.136$. Here, $ n_{\mathrm{e}} $ is the electron number density, and $ m_\mathrm{H} $ is the proton mass. With these quantities, the scale length can be computed as:

\begin{equation}
    a_\mathrm{g} = \frac{3 M_\mathrm{g}}{4 \pi \rho_0} ~.
    \label{eq:ag}
\end{equation}

Further details about the numerical implementation, including the velocity and temperature distributions, can be found in \citet{Machado13}. The initial conditions for NGC~5098/5096 were generated with the following numbers of particles for each of the  galaxy groups: $N_\mathrm{DM} = 10^6$ for the dark matter and $N_\mathrm{gas} = 10^6$ for the gas, using the \textsc{clustep}\footnote{\url{https://github.com/elvismello/clustep}} code \citep{Ruggiero17}. The specific parameters adopted for each galaxy group in the simulations are listed in Table\,\ref{parameters_table}.

\begin{table}
\centering
\caption{Initial parameters for the simulated galaxy groups. Columns list the virial mass $M_{200}$, concentration $c$, Hernquist scale length $a_h$, total gas mass $M_\mathrm{gas}$, and Dehnen scale length $a_g$ for NGC~5098 and NGC~5096.}
\begin{adjustbox}{width=1\columnwidth}
\begin{tabular}{cccccc}
\toprule
& $M_{200}$ ($10^{13}$ M$_\odot$) & $c$ & $a_\mathrm{h}$ (kpc) & $M_\mathrm{gas}$ ($10^{12}$ M$_\odot$) & $a_\mathrm{g}$ (kpc) \\
\midrule
NGC 5098 & $3.93$ & $5.2$ & $186$ & $6.34$ & $221$ \\
NGC 5096 & $0.17$ & $6.9$ & $54$  & $0.25$ & $76$ \\
\bottomrule
\end{tabular}
\end{adjustbox}
\label{parameters_table}
\end{table}

\subsection{Orbital parameters}

\begin{figure}
\begin{center}
\includegraphics[width=0.5\columnwidth]{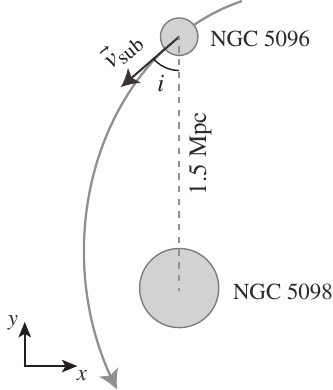}
\end{center}
\caption{Schematic representation of the collision in the orbital plane. Here, $v_{\mathrm{sub}}$ denotes the initial velocity of NGC~5096, and $i$ is the angle of incidence between the line joining the two galaxy groups and the direction of the initial velocity vector.}
\label{squeme}
\end{figure}

To investigate the impact of varying individual parameters in our models on the outcome of the simulations, all simulations were initialized with the same initial separation. Since the sum of the virial radii (assuming equal to $R_{200}$) of NGC~5098 and NGC~5096 is $0.93$\,Mpc, we adopted an initial distance of $d = 1.5$\,Mpc, which exceeds the combined virial radii and prevents significant overlap of dark matter and gas particles at the start of the simulations, $t = 0$. By fixing $d$, and consequently the initial gravitational potential energy, we ensure that any differences in the dynamical evolution arise from changes in other initial parameters, such as the initial infall velocity and the angle of incidence. A schematic description of the orbital plane is shown in Fig.~\ref{squeme}, where $\lvert \vec{v}_{\mathrm{sub}} \rvert = v_{\mathrm{inf}}$ is the initial infall velocity and $i$ is the angle of incidence.

To identify the best-fitting model, we ran several simulations, varying the initial conditions as described above. In this work, we present only the five models that we consider essential for understanding the dynamics of the NGC~5098/5096 system. These models are crucial for elucidating the development of edge-like features and the sloshing morphology. The corresponding dynamical parameters are listed in Table~\ref{initial_cond_table}.

\begin{table}
\centering
\caption{Dynamical parameters adopted for the initial conditions of each simulation model, including the initial separation $d$, infall velocity $v_{\mathrm{inf}}$, and angle of incidence $i$.}
\begin{adjustbox}{width=0.65\columnwidth}
\begin{tabular}{cccc}
\toprule
Scenario & $d$ [Mpc] & $v_\mathrm{inf}$ [km\,s$^{-1}$] & $i$ [$^{\circ}$] \\
\midrule
1 & $1.5$ & $650$ & $5$ \\
2 & $1.5$ & $650$ & $0$ \\
3 & $1.5$ & $650$ & $10$ \\
4 & $1.5$ & $950$ & $5$ \\
5 & $1.5$ & $350$ & $5$ \\
\bottomrule
\end{tabular}
\end{adjustbox}
\label{initial_cond_table}
\end{table}

\subsection{X-ray data}

X-ray analyses of the system NGC~5098/5096 have been conducted by \citet{Xue04, Mahdavi05,Randall09}, and \citet{Sun09}. In this work, we compare our simulation results to the observed X-ray emission in order to generate a comprehensive model that explains the dynamical state of NGC~5098/5096. To this end, we downloaded the three publicly available \textit{Chandra} observations with ObsID\,2231 (PI Fukazawa; 11\,ks), ObsID\,3352 (PI Ueda; 2.66\,ks), and ObsID\,6941 (PI Buote; 39.13\,ks). These datasets were used to produce exposure-map-corrected images. For the construction of the radial surface brightness profile, we utilized only the two longer exposures (11\,ks and 39.13\,ks). 

We reprocessed the \textit{Chandra} observations following standard procedures using the CIAO software package and the calibration dataset CALDB provided by the Chandra X-ray Center (CXC)\footnote{\url{https://cxc.harvard.edu/ciao/}}. The script \texttt{chandra\_repro} was used with default parameters to reprocess each observation individually. This task handles the creation of a bad pixel file, a redistribution matrix file (.rmf), and an ancillary response file (.arf), using the most up-to-date calibration files while taking the observation date into account. It then generates a new Type II PHA (pulse height amplitude) file, the event file that we will use in subsequent analysis.

We then combined the reprocessed event files using the \texttt{merge\_obs} script to generate a single, broad-band (0.5–7.0\,keV), exposure-map-corrected image. The exposure pointings are all different, so this task effectively builds a mosaic image, taking into account the different detectors and their response files. This final image was used as the basis for extracting radial surface brightness profiles in the northeastern and southwestern sectors of the NGC~5098 group. We did not subtract the background, since we will only use surface brightness ratios (we assume that after the exposure-map correction the background is flat in the area of the galaxy group).

\section{Results}
\label{results}

In this work, we present a numerical model that indicates the NGC~5096 galaxy group as the perturber. Our best model reproduces the northeast `tail' previously noted by \cite{Gastaldello07}, which can also be interpreted as the sloshing mentioned in \cite{Randall09}.

In order to enable a more direct comparison between our simulation results and the observed X-ray emission, we generated realistic \textit{Chandra} X-ray mock using the \texttt{pyXSIM} library\footnote{\url{https://hea-www.cfa.harvard.edu/~jzuhone/pyxsim/}} \citep{ZuHone16}. \texttt{pyXSIM} implements the PHOX algorithm \citep{Biffi12}, which simulates photon lists from astrophysical plasma by convolving the simulated gas radiative properties with instrumental responses. For each of our best-fitting and comparison models, we produced synthetic \textit{Chandra} images in the 0.5 -- 7.0\,keV band, adopting an exposure time of $62.79$\,ks, comparable to the observed exposure of NGC~5098. We fixed the metallicity at $0.3\,Z_\odot$ and assumed a hydrogen column density of $n_\mathrm{H} = 1.34 \times 10^{20}$\,cm$^{-2}$ \citep{HI4PI16} in the direction of NGC~5098. The resulting corrected mock images were used to extract surface brightness profiles and to visually compare sloshing features, providing a consistent framework for evaluating the fidelity of our dynamical models.

\begin{figure}
\begin{center}
\includegraphics[width=\columnwidth]{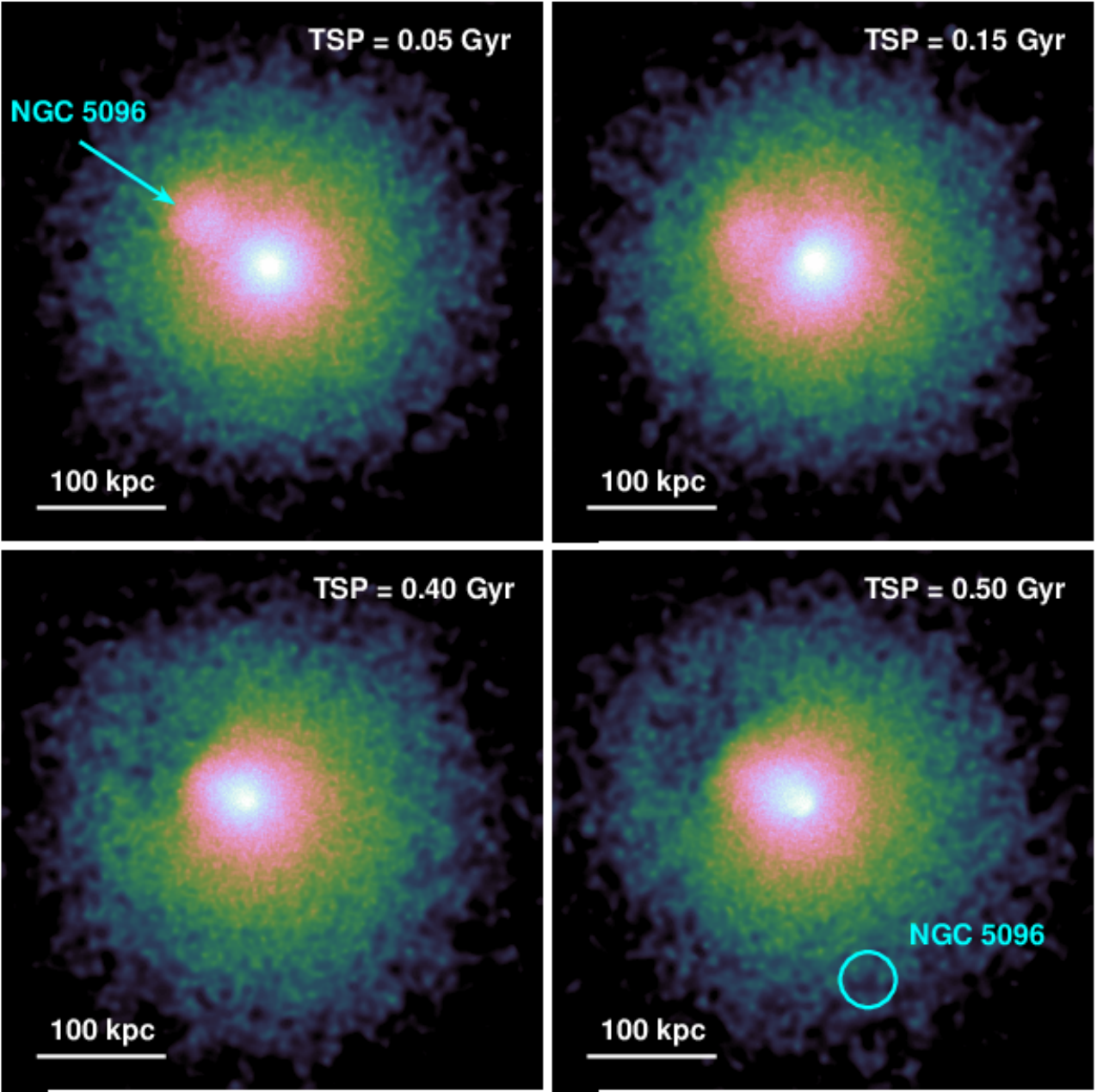}
\end{center}
\caption{Time evolution of the corrected mock \textit{Chandra} X-ray emissivity in the 0.5 -- 7.0\,keV band for the interaction between NGC~5098 and NGC~5096. TSP refers to the time since pericentre passage. In the final frame, the circle indicates the position of the NGC~5096 galaxy group.}
\label{NGC5098_time_evolution}
\end{figure}

The time evolution of the corrected mock \textit{Chandra} X-ray emissivity in the 0.5 -- 7.0\,keV band throughout the interaction is presented in Fig.~\ref{NGC5098_time_evolution}, with the last frame corresponding to the current dynamical stage of NGC~5098. The projected (three-dimensional) pericentric distance is approximately $30$\,kpc ($50$\,kpc), and the time since pericentric (TSP) passage is estimated to be TSP\,$=0.50$\,Gyr. Due to the relatively small impact parameter, the recent passage of the NGC~5096 subgroup triggered gas sloshing in the core of the main group, producing features that are consistent with those observed in the X-ray data. The simulation is shown with an inclination of $80^{\circ}$ between the orbital plane and the plane of the sky. This high inclination is required to reproduce the observed line-of-sight relative velocity between the galaxy groups. In our numerical experiment, NGC~5096 passes to the east relative to the centre of NGC~5098.

In the first panel of the time evolution of the emissivity shown in Fig.~\ref{NGC5098_time_evolution}, it is possible to observe NGC~5096 approaching from the northeast. After the pericentric passage, in the second panel, the gas associated with NGC~5096 begins to become more diffuse. At $\mathrm{TSP} = 0.40$\,Gyr, the northeastern edge starts to form, and the gas from NGC~5096 becomes almost indistinguishable from the IGM of NGC~5098. At $\mathrm{TSP} = 0.50$\,Gyr, there is no emissivity peak associated with NGC~5096, as almost all of its gas has already been stripped by ram pressure effects. This result agrees with observations, in which NGC~5096 shows only a faint gas peak. The gas originally belonging to NGC~5096 was dispersed into the IGM after the collision with NGC~5098. Since galaxies follow the dark matter distribution, in our analysis the position of NGC~5096 is given by the dark matter peak.

The edge-like discontinuities are formed immediately after the closest approach of the subgroup. As the perturber passes by, its gravitational influence displaces the cool, dense gas at the centre of NGC~5098 from the minimum of the gravitational potential and transfers angular momentum. This displacement triggers a characteristic sloshing motion of the intragroup gas, which manifests as sharp surface brightness discontinuities. These features initially emerge in the central regions and gradually propagate toward larger radii as the system continues to evolve dynamically. This process can be noted in the northeast edge region between $\mathrm{TSP} = 0.40$\,Gyr and $\mathrm{TSP} = 0.50$\,Gyr, as shown in Fig.~\ref{NGC5098_time_evolution}.

\begin{figure}
\begin{center}
\includegraphics[width=\columnwidth]{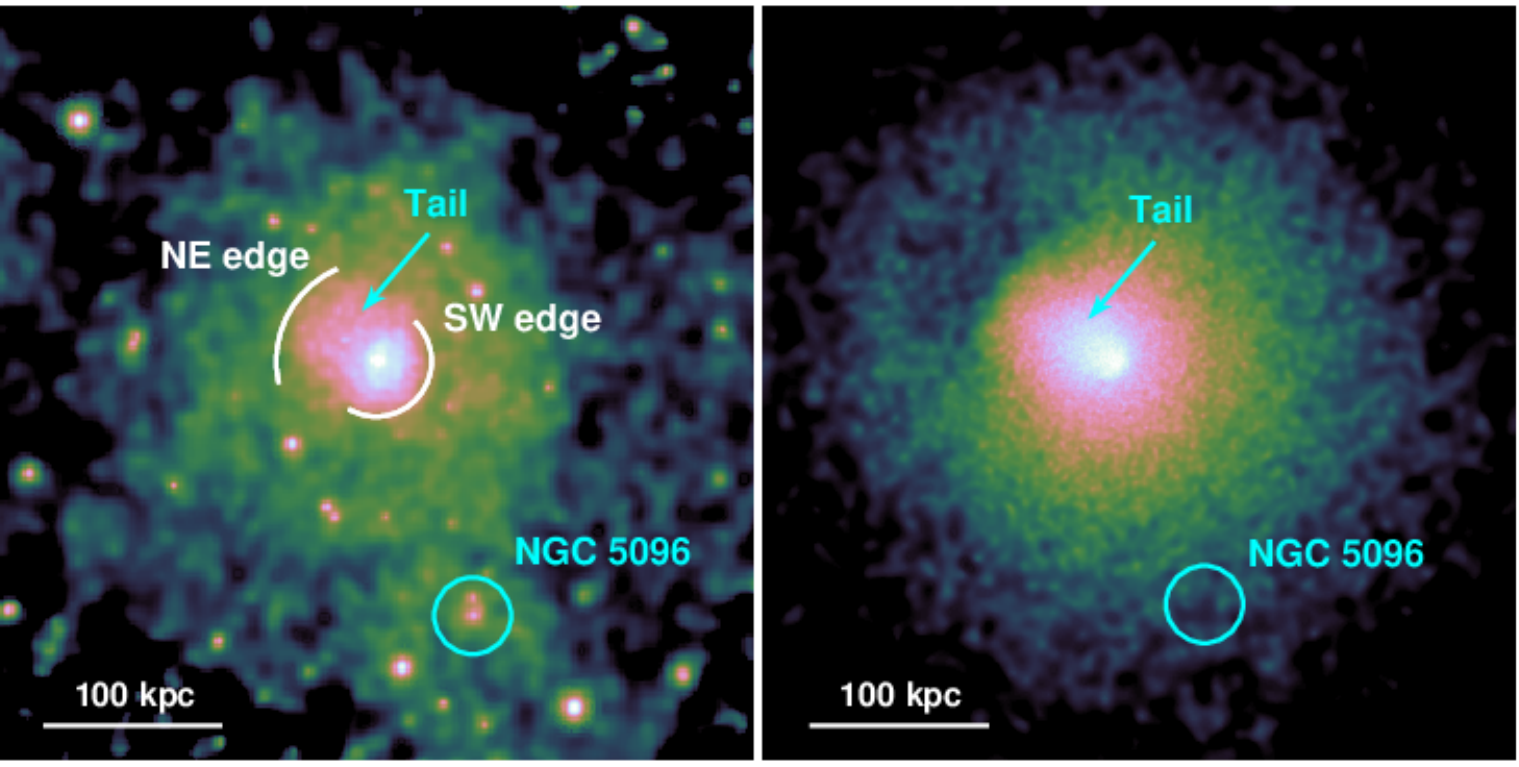}
\end{center}
\caption{Left: Smoothed, exposure-map corrected image of the combined \textit{Chandra} observations in the 0.5--7.0~keV band. The surface brightness edges identified by \citet{Randall09} are marked in white. Right: Smoothed, corrected mock \textit{Chandra} image in the same energy band, generated from the best-matching snapshot of our simulation. The simulated image successfully reproduces the main morphological features observed in the X-ray data, including the northeastern edge-like discontinuity associated with gas sloshing.}
\label{obs_mock_comp}
\end{figure}

For both the \textit{Chandra} and simulated data, we generated the exposure map in the broad band 0.5--7.0~keV. In Fig.~\ref{obs_mock_comp}, we compare the \textit{Chandra} image, corrected by the exposure map, with the corrected mock \textit{Chandra} image produced from the best-matching snapshot of our simulation. Each image has been smoothed for display purposes only. In this visual comparison, we note only a faint emission peak associated with NGC~5096 in the observations. In our simulated model, the gas originally belonging to NGC~5096 was almost completely stripped by the ram pressure effect, for this reason, no emissivity peak is present within the cyan circle. In the southwest sector, \citet{Randall09} identified an edge region corresponding to a discontinuity in the brightness profile, which we highlighted with a semicircular white arc in the left panel of Fig.~\ref{obs_mock_comp}, labeled as SW Edge. We do not see this discontinuity in the corrected mock image. To the northeast, we observe the presence of a tail associated with the sloshing phenomenon, visible in both the observations and the simulation. This tail, labeled in Fig.~\ref{obs_mock_comp}, appears spontaneously as a result of the interaction between NGC~5098 and NGC~5096 after the pericentric passage. The discontinuity in the northeast sector reported by \citet{Randall09} was also highlighted with a semicircular line in the exposure-corrected \textit{Chandra} image. In the corrected mock image, we can also see a discontinuity in that sector, although it appears at larger radii.

The comparison between the observed and simulated surface brightness maps, shown in Fig.~\ref{obs_mock_comp}, indicates that the exposure-corrected mock image reproduces the main morphological features seen in the X-ray data, including the X-ray excess in the northeastern region and the edge-like discontinuities to the northeast. 

\begin{figure}
\begin{center}
\includegraphics[width=\columnwidth]{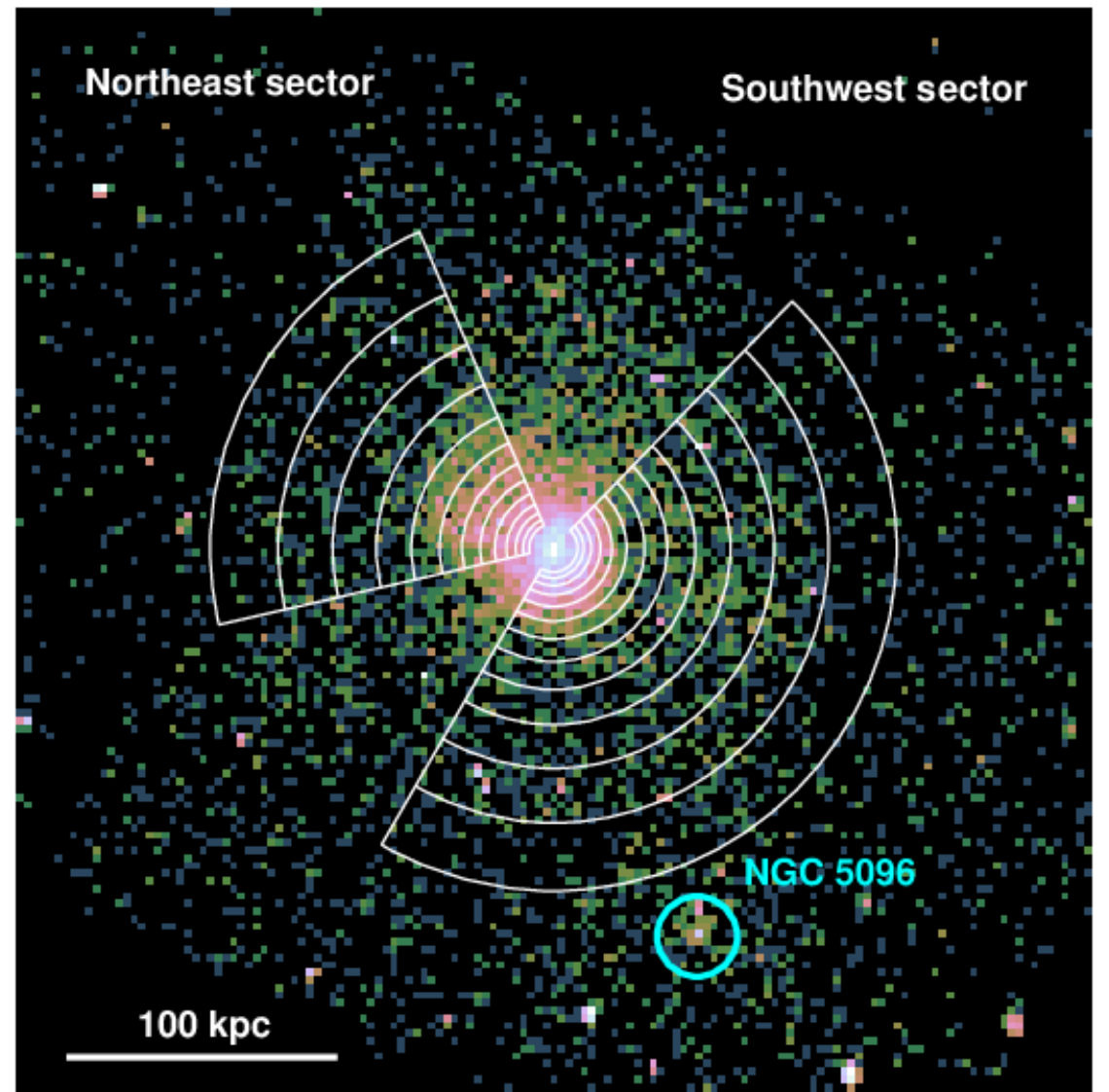}
\end{center}
\caption{Regions used to measure the X-ray surface brightness overlaid on the exposure-corrected \textit{Chandra} image. Each sector is centred on NGC~5098a and subdivided into 12 bins with logarithmic radial spacing. Source regions were masked out to extract only the diffuse gas surface brightness radial profile.}
\label{sectorsRegion}
\end{figure}

We also compare the surface brightness profiles from the X-ray observations with those derived from our simulated mock \textit{Chandra} image in the northeastern and southwestern sectors. By analyzing the X-ray surface brightness of NGC~5098 from the combined \textit{Chandra} observations, edge-like discontinuities were identified and can be associated with gas sloshing. To compare our simulation results with the observations, we defined a sector similar, though not identical, to that used in \citet{Randall09}. As shown in Fig.~\ref{sectorsRegion}, we defined one sector to the northeast and another to the southwest, each subdivided into 12 bins. For the observations, the sectors were centred on NGC~5098a, which coincides with the central emission peak. In the simulations, we centred the sectors on the simulated emission peak as well, to enable a more consistent comparison. The ratio was computed between the excess in each semicircular annulus and the average surface brightness of the respective sector. This same procedure was applied to both the observed \textit{Chandra} data and the simulated mock images.

\begin{figure}
\begin{center}
\includegraphics[width=\columnwidth]{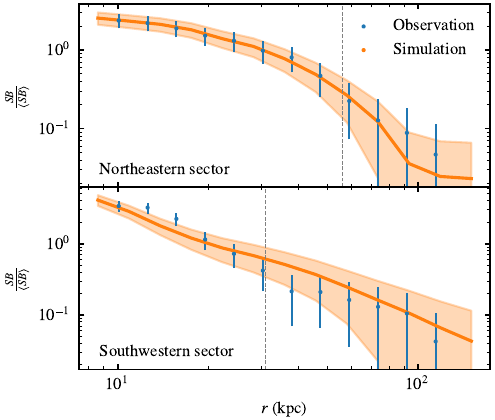}
\end{center}
\caption{Top panel: Surface brightness profile extracted from the northeastern sector for both the X-ray observations and the corrected mock \textit{Chandra} image. The vertical dashed line indicates the position of the surface brightness discontinuity reported by \citet{Randall09} at $r = 56$\,kpc, as highlighted in Fig.~\ref{obs_mock_comp}. The same extraction sector is applied over an extended radial range. Bottom panel: Same as the top panel, but for the southwestern sector. In this case, the vertical dashed line marks a radius of $r = 31$\,kpc. The orange shaded area indicates the $1 \sigma$ confidence interval.}
\label{NGC5098_flux_profile}
\end{figure}

As shown in Fig.~\ref{NGC5098_flux_profile}, the profile in the northeastern sector exhibits good agreement between the simulation and the observations, with the mean value of the simulated surface brightness lying within the observational uncertainties across the radial range. This agreement demonstrates that our model accurately reproduces the main features associated with the gas distribution, including the sloshing, observed in NGC~5098. In our simulation, the northeastern edge is detected at a radius of 93\,kpc, which is larger than the previously reported value. In the X-ray data, we observe only a slight decrease in the mean surface brightness at $r \sim 56$\,kpc. This result, derived from our simulated model, suggests that the sloshing northeastern edge in NGC~5098 may actually lie at a larger radius than the one detected by \citet{Randall09}.

For the southwestern sector, the mean value of our numerical model does not fully reproduce the observed surface brightness profile at certain radii. In the central region, this difference can be attributed to the presence of an AGN at the centre of NGC~5098. AGN jets inject energy into the IGM, heating the surrounding gas and inflating cavities or bubbles that appear as localized depressions in the X-ray surface brightness. These bubble regions are located to the north and southeast of the AGN. In the NGC~5098 group, these regions also exhibit associated radio emission that fills the X-ray cavities \citep[see][for further details]{Randall09}. When the confidence interval of the numerical model is taken into account, our best-fitting scenario remains consistent with the X-ray data, even in the absence of AGN feedback in the simulation.

Since we simulate the gravitational hydrodynamical interaction of a binary collision between two galaxy groups without including galaxies or feedback processes, our model cannot capture possible AGN-driven features in the central region. Additionally, the model only recovers the edge feature reported at $r = 31$\,kpc when the confidence interval is taken into account (see Fig.~\ref{NGC5098_flux_profile}). This surface brightness drop could be associated with older AGN bubbles, as suggested by the cavities visible near the southwestern edge in figure~3 of \citet{Randall09}. Alternatively, a substructure with a galaxy-scale mass could also induce a discontinuity in that region of the group. Although there is a region between $r = 20$\,kpc and $r = 30$\,kpc where the surface brightness decreases slightly more gradually, this variation is not sufficient to be classified as an edge-like feature.

Finally, from our best-fitting model, we obtained a line-of-sight relative velocity of 716\,km\,s$^{-1}$, which is comparable to the $\sim700$\,km\,s$^{-1}$ estimated from the observations, and a projected distance of 160\,kpc, which is very close to the observed value of 155\,kpc, further supporting the consistency of our model with the observational data.

\section{Discussion}

In the following, we discuss the impact of variations in key parameters, such as the initial infall velocity and the angle of incidence, which are directly related to the impact parameter of the NGC~5098/5096 system. We also explore the possibility that NGC~5098b may be responsible for the observed sloshing, and we compare the distribution of dark matter with the ICL to examine the alignment between these two components.

\subsection{Exploring the initial‐condition parameter space}

We performed a systematic exploration of the initial-condition parameter space to obtain a simulated model capable of explaining the dynamical scenario of NGC~5098. Our goal is to reproduce the sloshing signatures and obtain a surface brightness profile consistent with the observations. To this end, we tested different collision scenarios by varying key parameters. In this section, we discuss some of the crucial parameters explored, which have a direct impact on the resulting numerical model.

To enable a consistent comparison between models, we initialized each simulation with the same separation of $d = 1.5$\,Mpc between the two galaxy groups, which exceeds the sum of their virial radii ($R_{200}$ of NGC~5098 plus $R_{200}$ of NGC~5096). Ensuring that the initial distance is larger than the combined virial radii prevents any premature gravitational overlap or tidal interaction at $t = 0$\,Gyr. By fixing this initial separation, all models start with the same gravitational potential energy, allowing us to isolate and compare the effects of varying infall velocity and impact parameter.

\begin{figure}
\begin{center}
\includegraphics[width=\columnwidth]{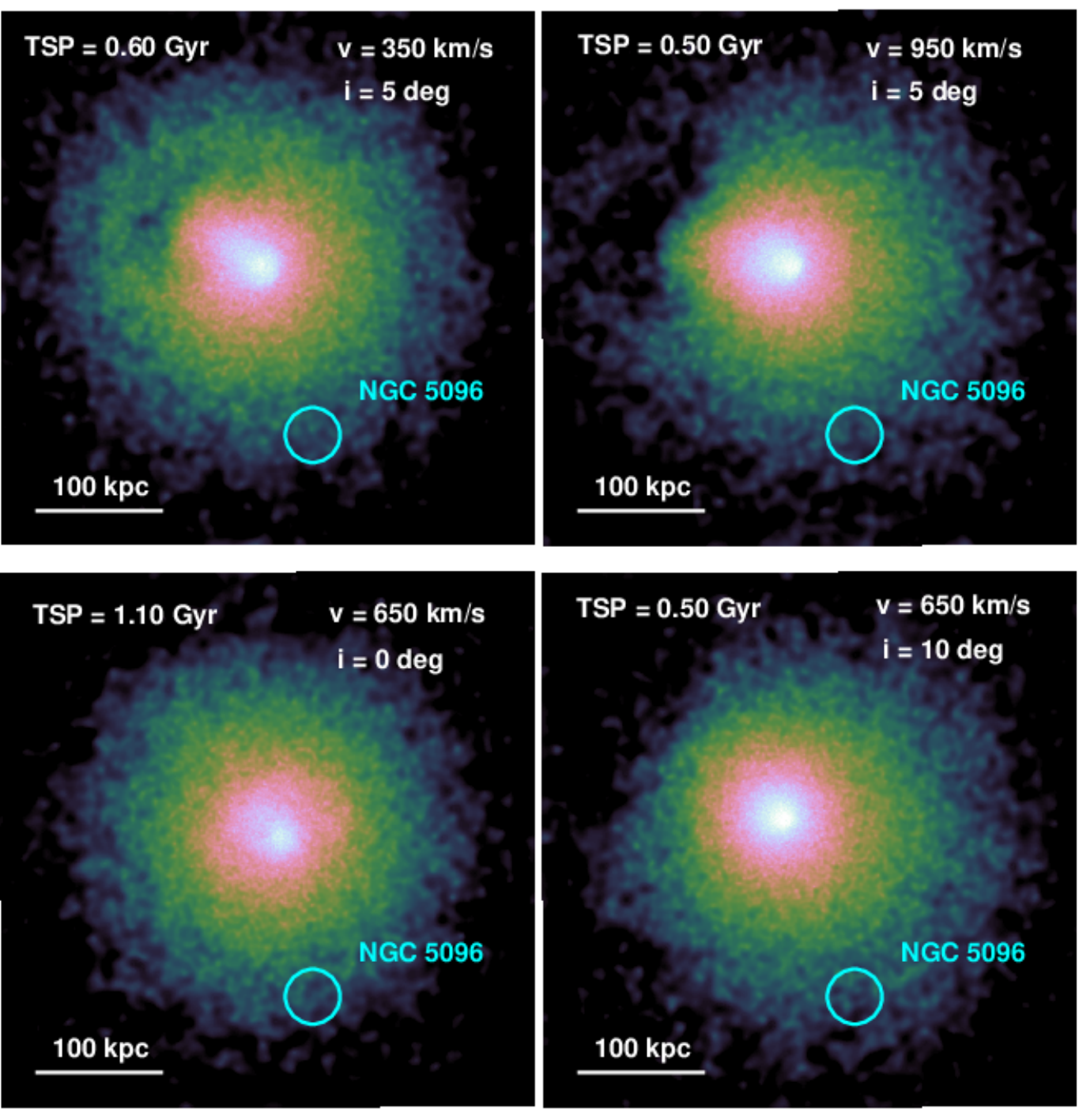}
\end{center}
\caption{Top panels: Comparison between simulations initialized with different infall velocities, keeping the initial angle of incidence of $5^\circ$. Bottom panels: Comparison between simulations with a fixed infall velocity of $650$\,km\,s$^{-1}$ and varying the angle of incidence. Our best-fitting model, shown in Fig.~\ref{obs_mock_comp}, corresponds to $v_{\mathrm{inf}} = 650$\,km\,s$^{-1}$ and $i = 5^\circ$. All models are projected assuming an inclination of $80^\circ$ between the orbital plane and the plane of the sky. The circle indicates the position of the NGC~5096 galaxy group.}
\label{comp_mock_all_models}
\end{figure}

In the top panels of Fig.~\ref{comp_mock_all_models}, we show two simulations at a fixed angle of incidence of $i = 5^\circ$, varying only the initial infall velocity. Specifically, one run uses $v_{\mathrm{inf}} = 350$\,km\,s$^{-1}$ and the other uses $v_{\mathrm{inf}} = 950$\,km\,s$^{-1}$. By keeping the separation fixed at $d = 1.5$\,Mpc and the angle of incidence constant, any differences in the resulting sloshing morphology can be directly attributed to the change in infall speed. This comparison highlights how sensitive the sloshing signature is to the galaxy groups relative velocity. Comparing both models, we find that lower initial infall velocities give the system more time to reach the expected projected separation, leading to a more developed sloshing pattern. For reference, a zero-energy orbit would have implied a relative velocity of 485\,km\,s$^{-1}$ in the initial conditions. This idealized estimate corresponds to the free-fall velocity of two point masses released from infinity. Thus, the different initial relative velocities in the simulations cover a range from 0.7 to 2 of the free-fall velocity. Due to the relatively low gas temperature, the sound speed in the intracluster medium, in the regions of interest, is in the range 350--500\,km\,s$^{-1}$, which implies supersonic speeds, even more so at the pericentric passage.

\begin{figure*}
\begin{center}
\includegraphics[width=\textwidth]{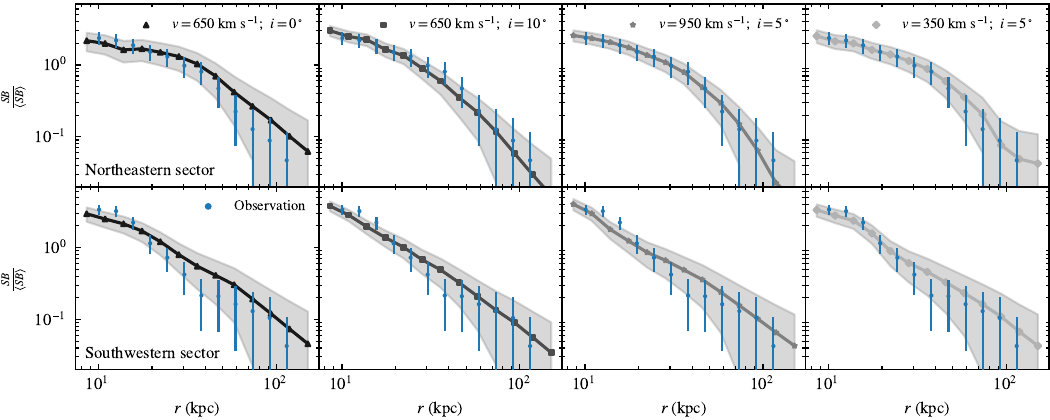}
\end{center}
\caption{Surface brightness profiles extracted from the northeastern (top panels) and southwestern (bottom panels) sectors, for both the X-ray observations and the simulated mock \textit{Chandra} images. Each column corresponds to a different collision scenario. In each panel, the blue lines represent the observed radial surface brightness profile, while the gray lines correspond to the different collision scenarios shown in Fig.~\ref{comp_mock_all_models}. The gray shaded area indicates the $1 \sigma$ confidence interval.}

\label{brightness_comp_mock_all_models}
\end{figure*}

Since we fixed the total potential energy of the system, the $v_{\mathrm{inf}} = 350$\,km\,s$^{-1}$ model begins with lower initial kinetic energy, resulting in a closer pericentric passage and an impact parameter $b = 63$\,kpc. In this case, a clear X-ray brightness discontinuity appears in the northeastern sector. By contrast, the $v_{\mathrm{inf}} = 950$\,km\,s$^{-1}$ model, which results in $b = 100$\,kpc, shows no such discontinuity. This difference is evident in Fig.~\ref{brightness_comp_mock_all_models}, where the surface brightness profiles are compared. Specifically, at a radius of $118$\,kpc, the northeast brightness profile in the $v_{\mathrm{inf}} = 350$\,km\,s$^{-1}$ model shows a discontinuity. In the higher-velocity scenario, however, the brightness declines smoothly at larger radii, without forming a sharp edge.

\begin{figure*}
\begin{center}
\includegraphics[width=\textwidth]{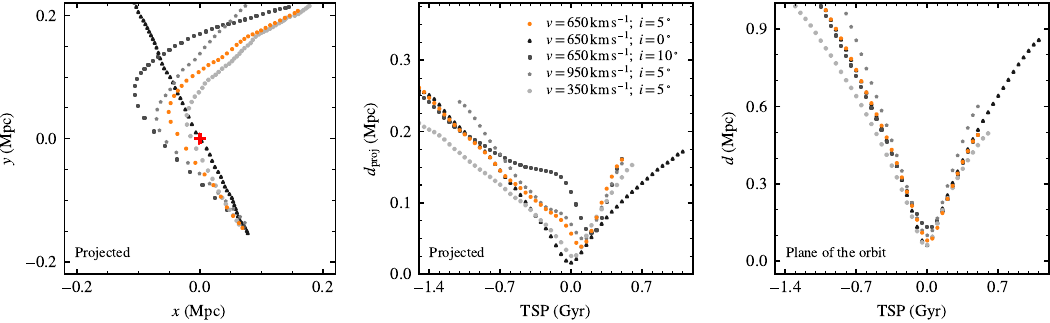}
\end{center}
\caption{Comparison of the NGC~5096 trajectory and distance evolution for different models. Left panel: Projected trajectories, where NGC~5098 is represented by the red cross at the origin. Middle panel: Projected separation as a function of TSP, measured from the plane of the sky configuration. Right panel: Three-dimensional separation as a function of TSP. The best-fitting model is shown with orange points, while the other models are displayed in shades of gray. All projections assume an inclination of $80^\circ$ between the orbital plane and the plane of the sky.}
\label{comp_trajectories}
\end{figure*}

In Fig.~\ref{comp_trajectories}, the left panel shows the trajectories of NGC~5096 for each scenario. Each collision scenario was run until the two groups reached a projected separation of approximately 150\,kpc; we also rotated the orbital plane to align NGC~5096 with its observed direction in the plane of the sky. We estimate the impact parameter from the minimum three-dimensional separation between NGC~5098 and NGC~5096, which occurs at $\mathrm{TSP} = 0$\,Gyr. The middle panel presents the projected separation as a function of TSP; due to line-of-sight effects, the minimum projected distance may occur at $\mathrm{TSP} \neq 0$\,Gyr.

Turning to the southwestern sector, the mean value of the lower-velocity scenario exhibits a significant mismatch in the surface brightness profile, whereas the higher-velocity scenario more closely follows the observational trend, similar to our best-fitting model. However, when the confidence interval is taken into account, both models remain consistent with the observations. Nevertheless, we discard both collision scenarios due to their inconsistency with the observed line-of-sight relative velocity. For the $v_{\mathrm{inf}} = 350$\,km\,s$^{-1}$ model, the line-of-sight relative velocity, after applying an inclination of $80^\circ$ between the orbital plane and the plane of the sky, is $v_{\mathrm{los}} = 474$\,km\,s$^{-1}$, which is lower than the $\sim700$\,km\,s$^{-1}$ inferred from observations. In contrast, the $v_{\mathrm{inf}} = 950$\,km\,s$^{-1}$ model yields $v_{\mathrm{los}} = 962$\,km\,s$^{-1}$, exceeding the expected value. Therefore, both scenarios fail to reproduce the kinematics of the NGC~5098 system. Our best-fitting model, with $v_{\mathrm{inf}} = 650$\,km\,s$^{-1}$, produces $v_{\mathrm{los}} = 716$\,km\,s$^{-1}$, which closely matches the observed velocity.

Fixing the initial potential and kinetic energies, we varied the angle of incidence to $0^\circ$ and $10^\circ$. For $i = 0^\circ$, the impact parameter is $b \approx 0$\,kpc, as shown by the dark gray points crossing the origin in the left panel of Fig.~\ref{comp_trajectories}, where NGC~5098 is located. In this configuration, the line-of-sight relative velocity is $v_{\mathrm{los}} = 522$\,km\,s$^{-1}$, which is lower than the observed value. For $i = 10^\circ$, the impact parameter increases to $b = 131$\,kpc, and we obtain $v_{\mathrm{los}} = 683$\,km\,s$^{-1}$. From a kinematic perspective, both configurations are able to reproduce the observed line-of-sight relative velocity of the NGC~5098 system.

When comparing the surface brightness profiles in Fig.~\ref{brightness_comp_mock_all_models}, the $i = 0^\circ$ model shows a brightness distribution that deviates significantly from the observations, even when the confidence interval is taken into account. Moreover, this model fails to reproduce the sloshing signature in the simulated mock \textit{Chandra} emissivity map, as shown in Fig.~\ref{comp_mock_all_models}. In contrast, the $i = 10^\circ$ model provides a fit similar to our best-fitting case; however, it does not produce an edge in the northeastern sector.
 
Overall, comparing our models, we find that collisions with smaller impact parameters are more effective at producing an edge-like feature in the northeastern sector. Models with initial infall velocities below $v_{\mathrm{inf}} = 350$\,km\,s$^{-1}$ fail to reproduce the observed velocity and edge position, even when large inclination angles between the line of sight and the orbital plane are applied. For the NGC~5098/5096 minor merger, with a mass ratio of 1:20, even head-on encounters with impact parameter $b \approx 0$ did not lead to any disruption of the cool core in the NGC~5098 galaxy group.

\subsection{NGC~5098b as an alternative perturber}

The sloshing arm detected by \citet{Randall09} appears as an X-ray emission excess to the north of NGC~5098b, spiraling outward in a clockwise pattern. This sloshing feature was attributed to the gravitational influence of NGC~5098b on NGC~5098a, the dominant galaxy in the group. In our orbital interpretation, a clockwise spiral requires the perturber to have approached from the northeast. The two galaxies are separated by a projected distance of $27$\,kpc, and spectroscopic redshifts indicate a line-of-sight relative velocity difference of $v_{\mathrm{los}} = 400$\,km\,s$^{-1}$. 

To reproduce these observational constraints, we initialized a collision scenario in which NGC~5098b starts at a distance of $1.5$\,Mpc from NGC~5098a, varying both the initial infall velocity and the angle of incidence. We then rotated the system to match the observed projected separation and line-of-sight velocity at the best-fitting moment. In this configuration, however, no tail formation was observed to the northeast of NGC~5098. The absence of a prominent sloshing feature suggests that NGC~5098b is unlikely to be the primary perturber of the NGC~5098/5096 system.

As discussed in section~\ref{results}, when the NGC~5096 galaxy group is instead considered as the main perturber, key features such as the tail morphology and the radial surface brightness profile are successfully reproduced. Therefore, although we cannot entirely rule out the possibility that NGC~5098b acted as the main perturber, our simulations indicate that this scenario is improbable.

\subsection{Comparison between dark matter distribution and intragroup light}

Some authors suggest \citep[e.g.][]{Montes19, Kluge21} that the ICL component typically aligns with host cluster properties, which makes it a potential dark matter tracer. \cite{Montes19} conducted morphological comparisons between ICL and dark matter, finding the ICL better than X-ray emission for tracing the total mass in central regions of Hubble Frontier Field clusters. Similarly, \cite{Alonso2020} using Cluster-EAGLE simulations confirm the correlation between ICL and total mass, despite radial density profiles showing mismatches. 

After obtaining a best-fitting collision scenario, we compare our simulation results with recent analyses of the ICL \citetalias{LimaNeto25} to check whether a correlation between dark matter and ICL also arises in our scenario. To this end, we compare the residual image of the simulated dark matter distribution with the observed ICL map.

\begin{figure}
\begin{center}
\includegraphics[width=\columnwidth]{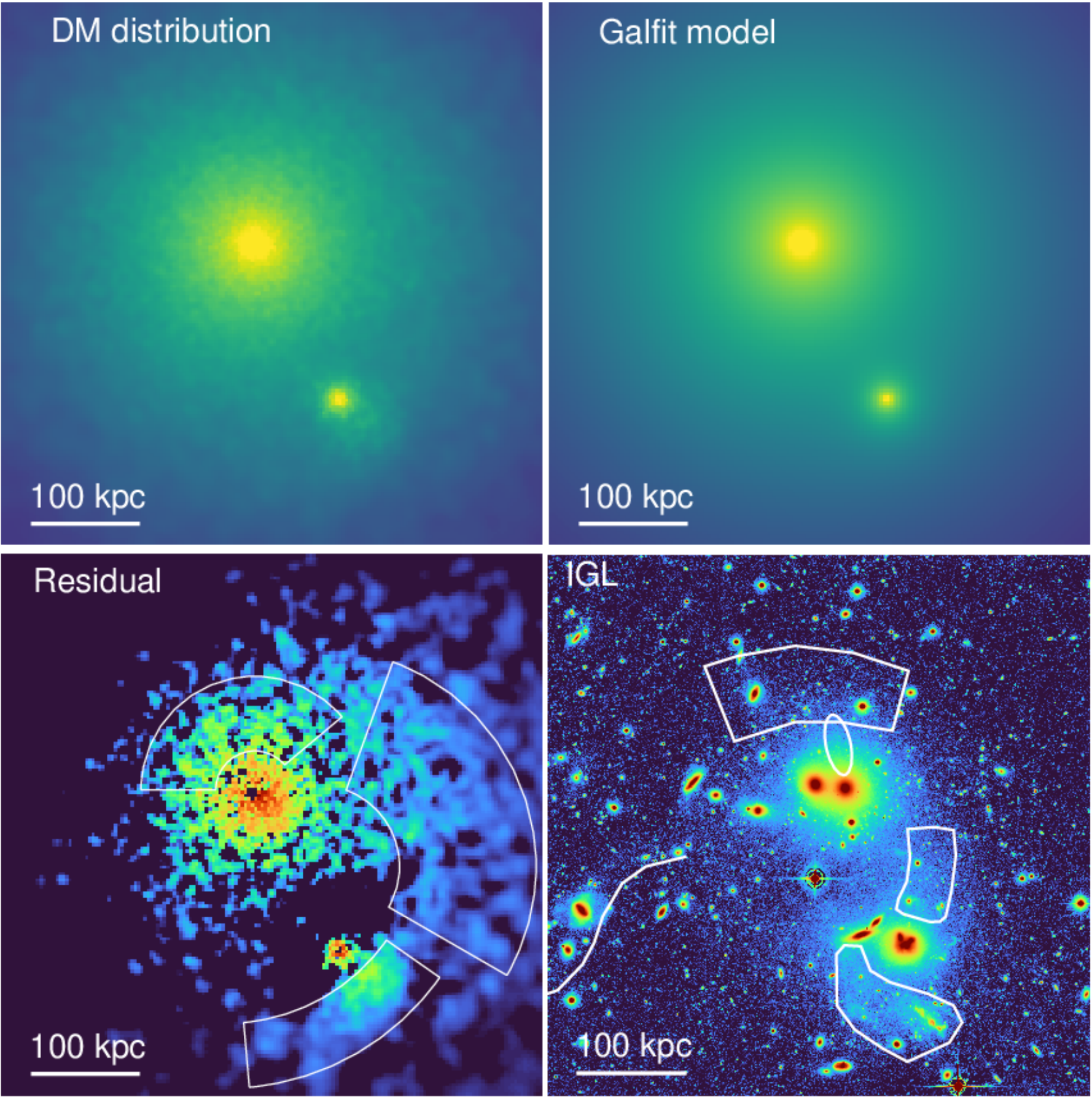}
\end{center}
\caption{The top-left panel shows our best-fitting model for the NGC~5098/5096 system, while the top-right panel presents a \textsc{GALFIT}-based model of both dark matter halos. The bottom panels compare the two: on the left, the residual map obtained by subtracting our best-fitting scenario from the \textsc{GALFIT} model; on the right, the $g$-band residual image after subtracting the stellar light models, highlighting the ICL within the white contours, adapted from \citetalias{LimaNeto25}.}
\label{ICL_dm_comp}
\end{figure}

The top-left panel of Fig.~\ref{ICL_dm_comp} shows the simulated dark matter distribution. We adapted the parameters of the \textsc{GALFIT} code \citep{Peng02} to fit a Nuker profile \citep{Lauer95} to the dark matter halos of both NGC~5098 and NGC~5096. The resulting \textsc{GALFIT} model is shown in the top-right panel. By subtracting this model from our best-fitting simulation, we obtain the residual image shown in the bottom-left panel. For comparison, the bottom-right panel presents the $g$-band residual image from \citetalias{LimaNeto25}, which highlights the ICL after stellar light subtraction.

We identify an excess of dark matter to the northeast of NGC~5098, highlighted by a semicircular annulus, which spatially coincides with an enhancement of the ICL in the same region. Similarly, another slight excess is found to the southwest of NGC~5098, also aligned with an ICL overdensity. Additionally, an excess of dark matter is found to the southeast of NGC~5096; however, in this case, we are not certain whether it is comparable to the ICL in the same region. This uncertainty arises because the diffuse light to the south of NGC~5096 may correspond to a tidal dwarf galaxy or to the remnant of a galaxy disrupted by tidal forces, as suggested by \citetalias{LimaNeto25}, and is not necessarily related to the NGC~5098/5096 interaction.

Comparing the ICL morphology with the dark matter density in Fig.~\ref{ICL_dm_comp}, we find a correlation between the two distributions, whose origin could be related to the NGC~5098/5096 interaction. However, the residual image derived from the dark matter distribution does not reproduce all the individual ICL features reported in \citetalias{LimaNeto25}.

\section{Conclusions}
\label{conclusions}

We have conducted tailored hydrodynamic $N$-body simulations of the NGC~5098/5096 system aiming to reproduce the observed gas sloshing morphology in the NGC~5098 galaxy group. By systematically exploring the initial condition parameter space, we identified the range of key parameters that best reproduce the observational data. Our results support the scenario in which NGC~5096 acts as the perturber, in agreement with our X-ray analysis. The main conclusions are summarized below:

\begin{enumerate}
    \item \textbf{NGC~5096 as the primary perturber.} Our simulations demonstrate that even a relatively low mass substructure can trigger gas sloshing in an intermediate mass galaxy group. Attempts to reproduce the sloshing signature by placing the perturber at the location of NGC~5098b (27\,kpc east) were unsuccessful. Instead, adopting NGC~5096 as the perturber yields our best-fitting model, which reproduces the projected separation in the plane of the sky ($d_{\mathrm{proj}} = 160$\,kpc), the line-of-sight relative velocity ($v_{\mathrm{los}} = 716$\,km\,s$^{-1}$), and the characteristic northeastern sloshing feature. Although this idealized model does not include AGN feedback, whose jets can create cavities that depress the X-ray surface brightness, our simulated brightness profile still matches the observations in the southwestern sector when the confidence interval is taken into account. 

    \item \textbf{Sensitivity to impact parameter.}  Our comparison of collision scenarios reveals that smaller impact parameters more effectively induce sloshing and generate edge-like discontinuities in the northeastern sector of NGC~5098. While a head-on collision ($b = 0$\,kpc) represents an extreme case, our best-fitting model requires an impact parameter of $b = 50$\,kpc to reproduce the key morphological features of the northeastern edge. In contrast, when we fix the initial infall velocity at $v_{\mathrm{inf}} = 650$\,km\,s$^{-1}$ and increase the angle of incidence to $i = 10^\circ$, the resulting impact parameter ($b = 131$\,kpc) is too large to produce a clear edge-like feature in the northeast; however, sloshing motions remain evident in the gas distribution.

    \item \textbf{Sensitivity to initial infall velocity.}  We tested three scenarios with varying initial infall velocities. Models with $v_{\mathrm{inf}} < 350$\,km\,s$^{-1}$ cannot reach the observed line-of-sight relative velocity of $\sim700$\,km\,s$^{-1}$, even when high orbital inclinations are applied. On the other hand, models with $v_{\mathrm{inf}} > 950$\,km\,s$^{-1}$ produce line-of-sight relative velocities that exceed the observed value. We also explored lower inclination angles; in those cases, the sloshing feature in the northeastern sector does not have sufficient time to develop before the system reaches the observed projected separation.

    \item \textbf{Intracluster light (ICL) comparison.} In our collision scenario, we find that the residual dark matter distribution roughly follows the same general morphology as observed in the ICL distribution obtained from observations. The diffuse light distribution in certain regions appears to be related to the NGC~5098/5096 interaction, although we were not able to reproduce all individual ICL features reported in previous analyses.
\end{enumerate}

Modelling galaxy groups in the literature is not common. This task is particularly challenging because even a single galaxy can perturb the IGM and non-gravitational processes such as AGN feedback can generate cavities that complicate the interpretation of the collision scenario. Our work demonstrates that the gravitational interaction with NGC~5096 provides a coherent explanation for the sloshing features in NGC~5098. Future efforts incorporating AGN feedback and more complex group environments will further refine our understanding of gas dynamics in galaxy groups.

\section*{ACKNOWLEDGEMENTS}

RPA acknowledges financial support from FAPESP under grant 2024/13224-9. GBLN is grateful for the financial support from CNPq under grant 314528/2023-7 and FAPESP under grant 2024/06400-5. REGM acknowledges support from CNPq, through grants 406908/2018-4 and 303425/2024-5, and from Fundação Araucária through grant PDI 346/2024 -- NAPI Fenômenos Extremos do Universo. F.D. acknowledges financial support from CNES.

\section*{Data Availability}

The data supporting this article will be shared upon reasonable request to the corresponding author.



\bibliographystyle{mnras}
\bibliography{NGC5098}




\bsp	
\label{lastpage}
\end{document}